\documentclass[aps, preprint, superscriptaddress]{revtex4-2}
\usepackage{graphicx}
\usepackage{makecell, multirow}
\usepackage{color, bm, soul, xcolor}
\usepackage{amsmath, amssymb, amsfonts}
\usepackage{gensymb}
\usepackage{array}
\usepackage{geometry}
\usepackage{braket}
\usepackage{diagbox}
\usepackage[colorlinks=true, linkcolor=blue, citecolor=blue, urlcolor=blue]{hyperref}
\newcommand{\eg}{{\em e.\,g.}}
\newcommand{\etal}{{\em et al.}}

\begin{document}

\title{Unexpected Density Functional Dependence of the Antipolar $\boldsymbol{Pbcn}$ Phase in HfO$\boldsymbol{_2}$}

\author{Di Fan}
\thanks{These two authors contributed equally.}
\affiliation{Fudan University, Shanghai 200433, China}
\affiliation{Department of Physics, School of Science, Westlake University, Hangzhou, Zhejiang 310030, China}

\author{Tianyuan Zhu}
\thanks{These two authors contributed equally.}
\affiliation{Department of Physics, School of Science, Westlake University, Hangzhou, Zhejiang 310030, China}
\affiliation{Institute of Natural Sciences, Westlake Institute for Advanced Study, Hangzhou, Zhejiang 310024, China}

\author{Shi Liu}
\email{liushi@westlake.edu.cn}
\affiliation{Department of Physics, School of Science, Westlake University, Hangzhou, Zhejiang 310030, China}
\affiliation{Institute of Natural Sciences, Westlake Institute for Advanced Study, Hangzhou, Zhejiang 310024, China}

\begin{abstract}
{The antipolar $Pbcn$ phase of HfO$_2$ has been suggested to play an important role in the phase transition and polarization switching mechanisms in ferroelectric hafnia. In this study, we perform a comprehensive benchmark of density functional theory (DFT) calculations and deep potential molecular dynamics (DPMD) simulations to investigate the thermodynamic stability and phase transition behavior of hafnia, with a particular focus on the relationship between the $Pbcn$ and ferroelectric $Pca2_1$ phases. Our results reveal significant discrepancies in the predicted stability of the $Pbcn$ phase relative to the $Pca2_1$ phase across different exchange-correlation functionals. Notably, the PBE and hybrid HSE06 functionals exhibit consistent trends, which diverge from the predictions of the PBEsol and SCAN functionals. For a given density functional, temperature-driven phase transitions predicted by DFT-based quasi-harmonic free energy calculations aligns with finite-temperature MD simulations using a deep potential trained on the same density functional. Specifically, the PBE functional predicts a transition from $Pca2_1$ to $Pbcn$ with increasing temperature, while PBEsol predicts a transition from $Pca2_1$ to $P4_2/nmc$. 
A particularly striking and reassuring finding is that under fixed mechanical boundary conditions defined by the ground-state structure of $Pca2_1$, all functionals predict consistent relative phase stabilities and comparable switching barriers as well as domain wall energies. These findings underscore the unique characteristics of the $Pbcn$ phase in influencing phase transitions and switching mechanisms in ferroelectric hafnia.
}
\end{abstract}

\maketitle

\clearpage

\section{INTRODUCTION}

Hafnia-based materials have emerged as a prominent research focus in the field of ferroelectric materials due to its exceptional properties and compatibility with silicon-based fabrication processes~\cite{Cheema20p478,Luo20p1391,Mikolajick20p1434}. Since the discovery of ferroelectricity in silicon-doped HfO$_2$ thin films in 2011~\cite{Boscke11p102903}, hafnia has shown tremendous potential for various applications, particularly in nonvolatile ferroelectric memory~\cite{Lee20p1343,Kim21peabe1341}. Unlike traditional perovskite ferroelectrics, where the spontaneous electrical polarization primarily arises from the displacement of transition metal cations, the ferroelectricity of HfO$_2$ originates from local displacements of oxygen anions~\cite{Zhu24p188}. Despite its promising properties, HfO$_2$ still faces considerable challenges in practical applications, including the metastability of its ferroelectric phases and excessively large coercive fields~\cite{Schroeder22p653,Noheda23p562}. The ferroelectric phase of HfO$_2$ is highly sensitive to external factors such as electric fields, temperature, and mechanical stress~\cite{Wei22p154101,Zhou22peadd5953}. The coexistence of multiple polar and nonpolar phases in as-grown films poses serious reliability issues for device performance~\cite{Zhu24p188}. The large switching barrier of HfO$_2$ leads to coercive fields that are 2--3 orders of magnitude higher than those in perovskite ferroelectrics, limiting the endurance and energy efficiency of hafnia-based devices~\cite{Mulaosmanovic21p9405215,Mueller13p93}. Therefore, understanding phase transitions among hafnia polymorphs and polarization switching mechanisms is important for overcoming these challenges and unlocking the full potential of hafnia-based ferroelectrics.

The $Pca2_1$ phase is widely recognized as the polar phase responsible for the ferroelectricity in HfO$_2$~\cite{Park15p1811,Sang15p162905,Huan14p064111}. Symmetry analysis reveals that the $Pca2_1$ phase is connected to an antipolar phase, $Pbcn$, through a single soft polar mode~\cite{Zhou22peadd5953}. When considering the $Pbcn$ phase as the reference structure, $Pca2_1$ HfO$_2$ can be interpreted as a proper ferroelectric, characterized by a symmetric double-well potential~\cite{Raeliarijaona23p094109}. This contrasts with interpretations that use the tetragonal ($T$) $P4_2/nmc$ phase as the reference, in which case the ferroelectricity would be classified as improper~\cite{Lee20p1343,Delodovici21p064405}. Further insight into the appropriate nonpolar reference phase is provided by the minimum energy pathway for the polarization reversal in the $Pca2_1$ unit cell, identified using the nudged elastic band (NEB) method based on density functional theory (DFT) calculations. These calculations indicate that the transition state adopts the $Pbcn$ phase, rather than the $T$ phase, during polarization switching~\cite{Zhou22peadd5953}. These results highlight the $Pbcn$ phase as a critical reference state for understanding phase transitions and polarization switching mechanisms in  hafnia.

The temperature-driven ferroelectric-paraelectric phase transition offers valuable insight into the parent phase of the $Pca2_1$ structure and the underlying nature of ferroelectricity. However, interpreting experimental results, such as the dielectric anomaly observed upon heating Hf$_{0.5}$Zr$_{0.5}$O$_2$~\cite{chen2020p100919,zhou24p101414,adkins20p142902}, is challenging due to the coexistence of multiple hafnia polymorphs in as-grown thin films. Based on mode analysis, Raeliarijaona and Cohen suggested that at elevated temperatures, atoms within the $Pbcn$ phase can migrate between up and down positions, ultimately leading to the tetragonal structure~\cite{Raeliarijaona23p094109}. Molecular dynamics (MD) simulations by Wu \etal~have provided additional insights into the critical role of the $Pbcn$ phase, revealing its significance as an intermediate state in temperature-driven phase transitions~\cite{Wu23p144102}. Its role, however, varies depending on material composition. Specifically, in Hf$_x$Zr$_{1-x}$O$_2$ solid solutions, higher hafnium content ($x=0.5, 0.75, 0.9, 1.0$) leads to a transition from the polar $Pca2_1$ phase to the $Pbcn$ phase with increasing temperature. In contrast, zirconium-rich compositions ($x=0, 0.25$) bypass the $Pbcn$ phase and directly transition to the nonpolar $T$ phase during heating~\cite{Wu23p144102}. Further studies on strain effects, such as those by Wei \etal, have demonstrated that applying a 3\% lattice tensile strain along the (001) and (010) planes of the orthorhombic HfO$_2$ structure induces a ferroelectric-to-paraelectric phase transition, shifting from the polar $Pca2_1$ phase to the antipolar $Pbcn$ phase~\cite{Wei22p154101}. Moreover, the local structure of the $Pbcn$ phase shows similarities to certain types of domain walls that separate oppositely polarized $Pca2_1$ domains, suggesting that the $Pbcn$ phase may play an important role in the formation and evolution of domain walls~\cite{Wu24parXiv,Zhu25p056802}. Together, these findings underscore the multifaceted role of the $Pbcn$ phase in hafnia-based ferroelectrics, particularly its critical function in phase transition mechanisms and polarization switching dynamics.

In this work, we perform comprehensive benchmark calculations to evaluate the lattice structures and thermodynamic stability of several representative crystalline phases of HfO$_2$, with a particular focus on the $Pbcn$ phase. Notably, we observe significant discrepancies in the predicted relative thermodynamic stability of the $Pbcn$ phase relative to the $Pca2_1$ phase across different exchange-correlation (XC) density functionals. 
For example, the Perdew-Burke-Ernzerhof (PBE) functional predicts the $Pbcn$ phase to be more stable than $Pca2_1$, whereas PBEsol predicts the opposite. 
We find that predictions using the PBE functional match those obtained from the supposedly more accurate Heyd-Scuseria-Ernzerhof (HSE) hybrid density functional~\cite{Heyd03p8207, Paier06p154709, Marsman08p064201}. Interestingly, phonon dispersion calculations confirm the dynamic stability of the $Pbcn$ phase under all tested XC functionals, despite the differences in relative thermodynamic stability. We further use MD simulations to investigate temperature-driven phase transitions using deep potential (DP) models trained on first-principles database generated using PBE and PBEsol functionals, respectively. We observe that MD simulations using PBE-based DP model predict the $Pbcn$ phase as an intermediate high-temperature phase, whereas the PBEsol-based DP model reveals a direct $Pca2_1 \rightarrow T$ transition with increasing temperature. These MD results are consistent with quasi-harmonic free-energy calculations using the respective XC functional employed to train the DP models. Importantly, when the lattice constants are fixed to the ground-state values of the $Pca2_1$ phase, different XC functionals yield consistent results for key physical properties, such as polarization switching barriers, strain effects, and domain wall energies. This consistency offers reassurance regarding the accuracy of previous investigations of HfO$_2$ using PBEsol and LDA~\cite{Hinuma17p094102, Jaffe05114107, Duncan16043501} by fixing  lattice constants, despite variations in the predicted thermodynamic stability of the $Pbcn$ phase across functionals.

\section{METHODS}

We investigate the properties of the $Pbcn$ phase and its role in temperature-driven phase transitions for HfO$_2$ starting from the $Pca2_1$ phase using a combination of DFT calculations and deep potential molecular dynamics (DPMD) simulations~\cite{Zhang18p143001,Zhang18p4441}. DFT calculations are performed using the Vienna \textit{ab initio} simulation package (\texttt{VASP})~\cite{Kresse96p11169,Kresse96p15} with the projector augmented wave (PAW) method~\cite{Blochl94p17953, Kresse99p1758}. A comprehensive evaluation of XC density functionals is carried out, including PBE~\cite{Perdew96p3865}, PBEsol~\cite{Perdew08p136406}, local density approximation (LDA)~\cite{Kohn65pA1133}, strongly constrained and appropriately normed (SCAN)~\cite{Sun15p036402}, and the hybrid functional HSE06~\cite{Heyd03p8207}. The plane-wave cutoff energy is set to 600 eV, and a $4\times4\times4$ Monkhorst-Pack (MP) grid~\cite{Monkhorst76p5188} is used for Brillouin zone sampling. Convergence tests ensure the reliability of these parameters, and all structures are fully optimized until the atomic forces converge to a threshold of 0.001 eV/\AA. To validate the robustness of our results, we additionally employ the \texttt{PWmat} package~\cite{Jia13p102,Jia13p9} to verify the ground-state structures using aforementioned XC functionals with the norm-conserving pseudopotentials (\texttt{NCPP-SG15-PBE} and \texttt{NCPP-SG15-LDA})~\cite{Hamann13p085117,schlipf15p36} and a cutoff energy of 70 Ry. The HSE06 hybrid functional uses a mixing parameter $\alpha=0.25$ and a screening parameter $\omega=0.20~$\AA$^{-1}$. The polarization reversal pathway is identified using the DFT-based NEB method~\cite{Sheppard12p074103}.

Phonon properties are analyzed using a $2\times2\times2$ supercell (96 atoms), with the interatomic force constants determined via density functional perturbation theory (DFPT)\cite{Baroni01p515}. Phonon dispersion relations and Gibbs free energy curves under the quasi-harmonic approximation (QHA) are computed using the \texttt{Phonopy} package\cite{Togo23p353001}, incorporating non-analytical corrections through Born effective charges (BEC).

The DP model is a deep neural network (DNN) that maps local atomic environment to the atomic energy. In previous work, we developed a DP model for HfO$_2$ using a database of DFT energies and atomic forces computed with PBE functional. This database contains 21,768 configurations, representing various phases of HfO$_2$ (space groups: $P2_1/c$, $Pbca$, $Pca2_1$, and $P4_2/nmc$). Further details on the construction of this database can be found in ref.~\cite{Wu21p024108}. Here, we construct a PBEsol-based DP model by recalculating energies and forces for the same configurations with the PBEsol functional. The model is trained using the \texttt{DeePMD-kit} package~\cite{Wang18p178}, retaining the same DNN architecture and training parameters as in the previous work.

The temperature-driven phase transitions are modeled by performing constant-temperature, constant-pressure ($NPT$) MD simulations using a $10 \times 10 \times 10$ supercell containing 12,000 atoms and a time step of 1 fs. Temperature and pressure are controlled using the Nos\'e–Hoover thermostat~\cite{Hoover85p1695} and the Parrinello-Rahman barostat, respectively, implemented in the \texttt{LAMMPS} software~\cite{Plimpton95p1}. For each temperature increment, the final configuration from a lower-temperature simulation is used as the starting point for the next higher-temperature run. At each temperature, the system undergoes an equilibration phase of 30 ps, followed by a production run of 70 ps.

\section{RESULTS AND DISCUSSION}

\subsection{Ground state properties}

\begin{figure}[t]
\includegraphics[width=0.9\textwidth]{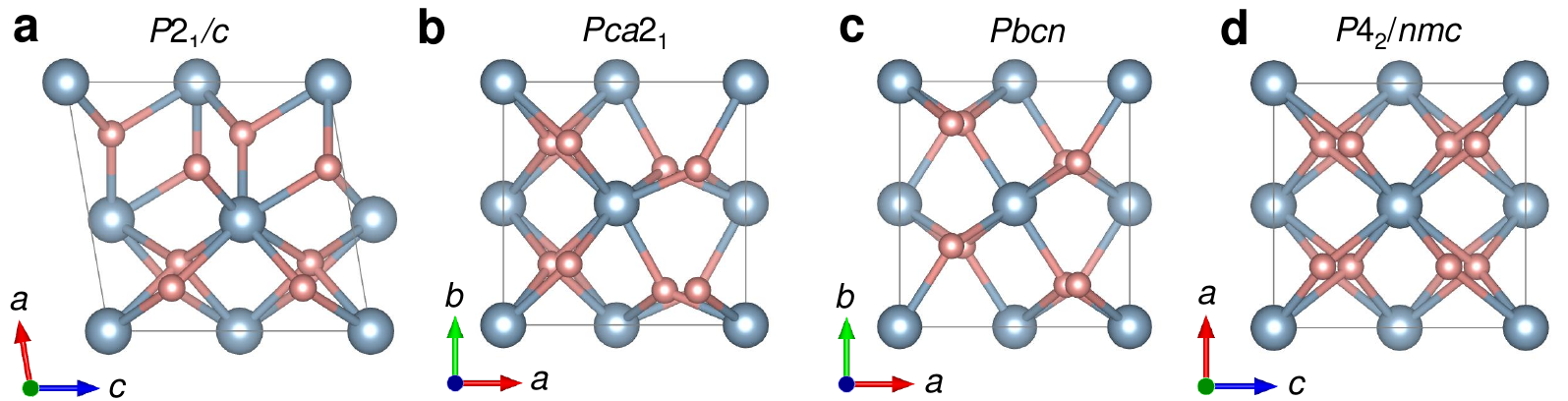}
\caption{Atomic structures of HfO$_2$ polymorphs. Hafnium (Hf) and oxygen (O) atoms are represented by blue and red spheres, respectively. (\textbf{a}) Monoclinic $P2_1/c$, (\textbf{b}) polar orthorhombic $Pca2_1$, (\textbf{c}) antipolar orthorhombic $Pbcn$, and (\textbf{d}) nonpolar tetragonal $P4_2/nmc$.}
\label{fig:structure}
\end{figure}

Despite the simple chemical composition, HfO$_2$ exhibits a rich array of polymorphic phase transitions. In this study, we focus on the $Pbcn$ phase and its stability relative to other polymorphs, including the monoclinic ($M$) $P2_1/c$, tetragonal ($T$) $P4_2/nmc$, and polar orthorhombic ($PO$) $Pca2_1$ phases, as illustrated in Fig.~\ref{fig:structure}. At room temperature and ambient pressure, the $M$ phase is the most stable phase of HfO$_2$. As the temperature increases, the $M$ phase gradually transforms into the $T$ phase and eventually transitions into the cubic ($C$) $Fm\bar{3}m$ phase at elevated temperatures. The ferroelectric properties observed in HfO$_2$-based thin films are commonly associated with the metastable $PO$ phase.

\begin{figure}[t]
\includegraphics[width=1.0\textwidth]{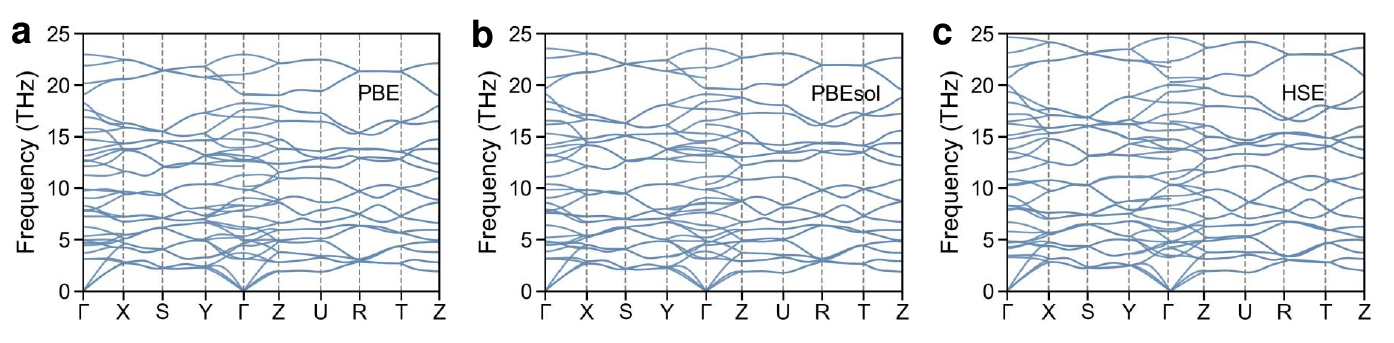}
\caption{Phonon spectra of the $Pbcn$ phase of HfO$_2$. Calculations are performed using (\textbf{a}) PBE, (\textbf{b}) PBEsol, and (\textbf{c}) HSE06 functionals, with non-analytical corrections included.}
\label{fig:phonon}
\end{figure}

We determine the lattice constants and relative energies of four polymorphs of HfO$_2$: $M$, $Pbcn$, $T$, and $PO$, using various XC functionals. The results, summarized in Table~\ref{tab:phase}, reveal general consistency in the relative energetic ordering obtained from \texttt{PWmat} and \texttt{VASP}, with calculated lattice constants also showing strong agreement between the two computational packages. However, discrepancies emerge in the relative thermodynamic stability of the $Pbcn$ phase depending on the XC functional. For the PBE and HSE06 functionals, the energy order is $M < Pbcn < PO < T$, with the $Pbcn$ phase being more stable than the $PO$ phase. In contrast, PBEsol, LDA, and SCAN all predict a markedly less stable $Pbcn$ phase, yielding a distinct energy order: $M < PO < Pbcn < T$. This difference in energetic order highlights the strong sensitivity of the $Pbcn$ phase's stability to the choice of XC functional. Despite the functional-dependent variations in relative stability, the computed phonon spectra for the $Pbcn$ phase using both PBE and PBEsol exhibit no imaginary frequencies across all wavevectors (Fig.~\ref{fig:phonon}), confirming its dynamic stability regardless of the functional employed. It is noted that we also calculate the relative energies of ZrO$_2$ polymorphs using both PBE and PBEsol. Although both functionals consistently predict the energy ordering as $PO < Pbcn$, they differ in the relative stability of the $Pbcn$ and $T$ phases, with the energy ordering between these two phases being reversed. 

\begin{figure}[t]
\includegraphics[width=0.9\textwidth]{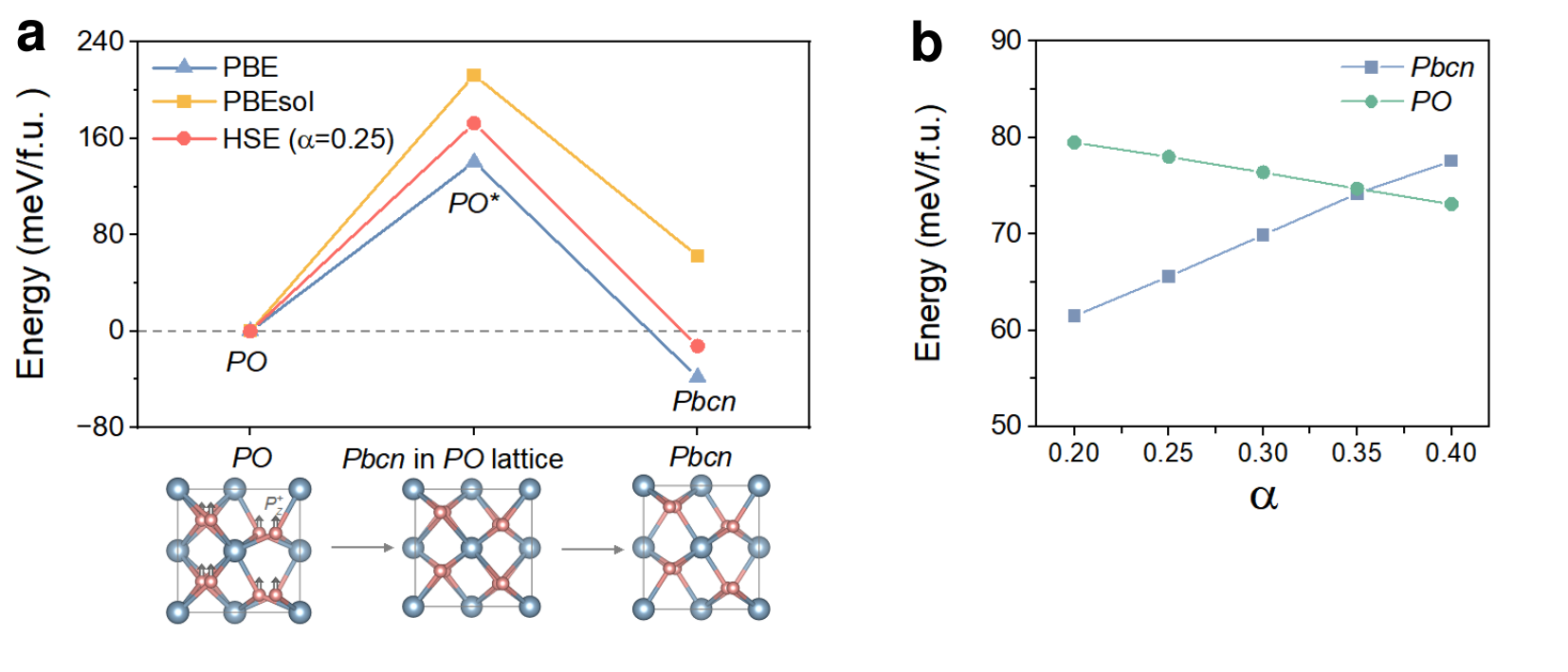}
\caption{(\textbf{a}) Energy evolution during the structural transition of $PO \rightarrow PO^* \rightarrow Pbcn$. The $PO^*$ phase is constructed by displacing oxygen atoms along the $P_z$ mode while keeping the lattice constants fixed to those of the $PO$ phase. Energy is referenced to the $PO$ phase ($E_{PO}$ = 0). (\textbf{b}) Variation in the relative energies of the $Pbcn$ and $PO$ phases as a function of $\alpha$ in HSE06.}
\label{fig:modecalculation}
\end{figure}

It is intriguing that, despite PBE and PBEsol predicting similar lattice constants for the $PO$ and $Pbcn$ phases, their predicted thermodynamic stability is reversed. To investigate this discrepancy, we calculate the energy of an intermediate structure that bridges the two phases. The structure, denoted as $PO^*$, is generated by displacing oxygen atoms along the $P_z$ mode in $PO$ until the mode amplitude reaches zero, while keeping the lattice parameters fixed. The $PO^*$ phase can be understood as a $Pbcn$-like phase  with the lattice constants of the $PO$ phase. As shown in Fig.~\ref{fig:modecalculation}a, all three functionals, PBE, PBEsol and HSE06 (with $\alpha=0.25$), predict a sharp energy increase when transitioning from $PO$ to $PO^*$: 140.6 meV/f.u.~for PBE and 212.7 meV/f.u.~for PBEsol, and 172.6 meV/f.u.~for HSE06. The subsequent transformation from $PO^*$ to $Pbcn$ mainly involves lattice expansion, which leads to a substantial energy decrease across all functionals. While PBE, PBEsol, and HSE06 exhibit a consistent overall trend, the extent of the energy reduction due to lattice expansion varies. Compared to PBEsol, both PBE and HSE06 predict greater stabilization of $Pbcn$, ultimately making it the lower-energy phase. This enhanced stability with increasing volume reflects PBE’s well-documented tendency to favor lower-density structures~\cite{Wahl08p104116,Wu06p235116,Perdew08p136406}.

Moreover, our HSE06 calculations with varying $\alpha$ values reveal that the relative stability between $PO$ and $Pbcn$ is sensitive to the exact exchange contribution. As shown in Fig.~\ref{fig:modecalculation}b, $Pbcn$ is more stable than $PO$ at lower $\alpha$, but as $\alpha$ increases, its energy gradually rises while $PO$ becomes increasingly stabilized. This trend reduces the energy gap between the two phases, ultimately leading to a stability reversal. Specifically, at $\alpha=0.35$, $Pbcn$ and $PO$ have nearly equal energies, and at $\alpha=0.4$, $PO$ becomes the more stable phase. Since enhanced HF exchange promotes electron localization and weakens the screening of long-range electrostatic interactions, the $\alpha$-driven energy reversal suggests that the stability of the antipolar $Pbcn$ phase is highly sensitive to the screening strength of antipolar distortions: weaker screening significantly destabilizes this phase.

\subsection{Temperature driven phase transition}

\begin{figure}[t]
\includegraphics[width=0.8\textwidth]{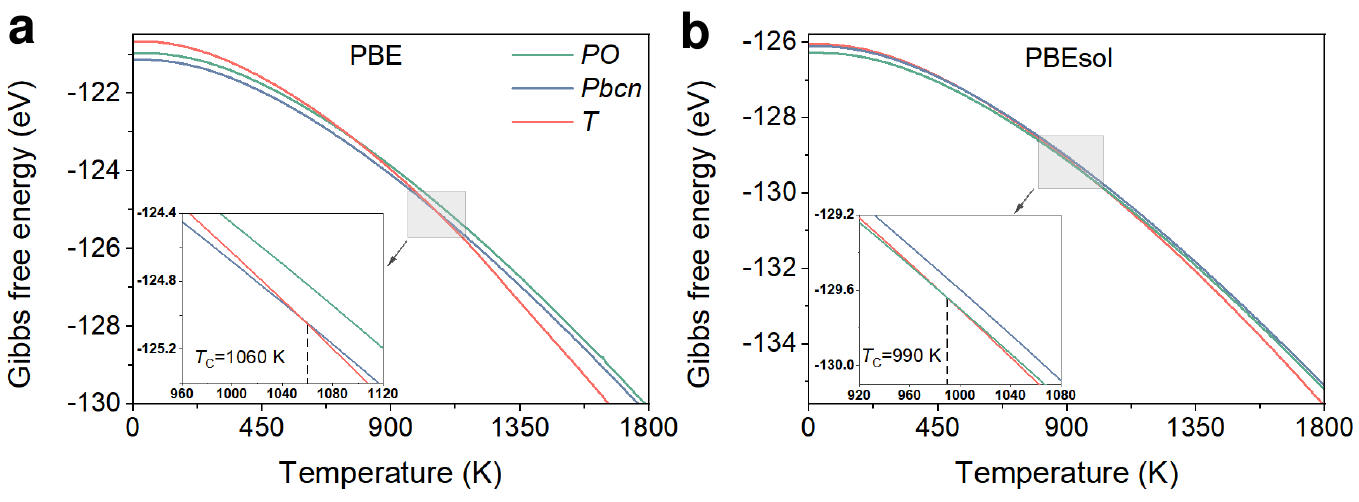}
\caption{Temperature-dependent Gibbs free energy of different HfO$_2$ polymorphs, calculated using (\textbf{a}) PBE and (\textbf{b}) PBEsol functionals based on QHA approximations.}
\label{fig:gibbs}
\end{figure}

To investigate the finite-temperature relative stability of the $Pbcn$, $PO$, and $T$ phases, we compute their temperature-dependent Gibbs free energy curves (Fig.~\ref{fig:gibbs}) using the QHA with both PBE and PBEsol functionals. As shown in Fig.~\ref{fig:gibbs}a, the PBE functional predicts that the $Pbcn$ phase maintains a consistently lower free energy than the $PO$ phase up to 1800 K. This implies that the $Pbcn$ phase will not spontaneously transition into the $PO$ phase within this temperature range. However, as the temperature increases, the free energy of the $T$ phase decreases more rapidly, eventually becoming the most stable phase at approximately 1100 K. Thus, QHA calculations based on the PBE functional predict a temperature-driven phase transition from $Pbcn$ to $T$ at around 1100~K. These results further suggest that if the system initially starts in the metastable $PO$ phase, increasing thermal fluctuations could drive the system out of the local free energy well, allowing it to evolve into the more stable $Pbcn$ phase. This observation is confirmed by our DPMD simulations (see further discussion below). In stark contrast to the PBE predictions, the PBEsol functional yields an entirely different scenario. It predicts that the $PO$ phase remains more stable than the $Pbcn$ phase throughout the entire temperature range up to 1800 K, with a phase transition from the $PO$ phase to the $T$ phase occurring above 950 K.

\begin{figure}[t]
\includegraphics[width=0.6\textwidth]{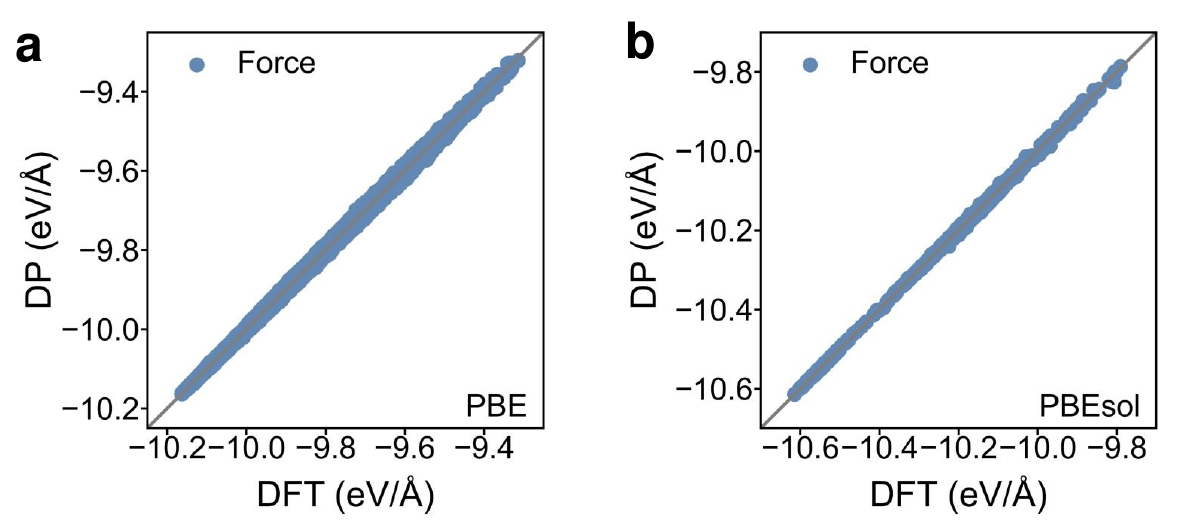}
\caption{Comparison of atomic forces predicted by the DP model with reference DFT results obtained using (\textbf{a}) PBE and (\textbf{b}) PBEsol, for configurations in the final training database.}
\label{fig:accuracy}
\end{figure}

Since the QHA assumes only small anharmonic effects, it may not accurately capture the behavior of systems near their phase transition temperatures. To address this limitation, we develop DP models, trained on a database of DFT-calculated energies and atomic forces, using PBE and PBEsol functionals separately. These DP models enable MD simulations that are capable of describing strongly anharmonic effects at finite temperatures. Figure~\ref{fig:accuracy} compares the atomic forces predicted by DFT and DP models for all structures in the training database. Both the PBE-based and PBEsol-based DP models exhibit excellent accuracy, achieving mean absolute errors (MAEs) of 3.5 meV/atom and 2.9 meV/atom, respectively. These results confirm the reliability of the DP models in faithfully reproducing the underlying DFT potential energy surfaces.

\begin{figure}[t]
\includegraphics[width=1.0\textwidth]{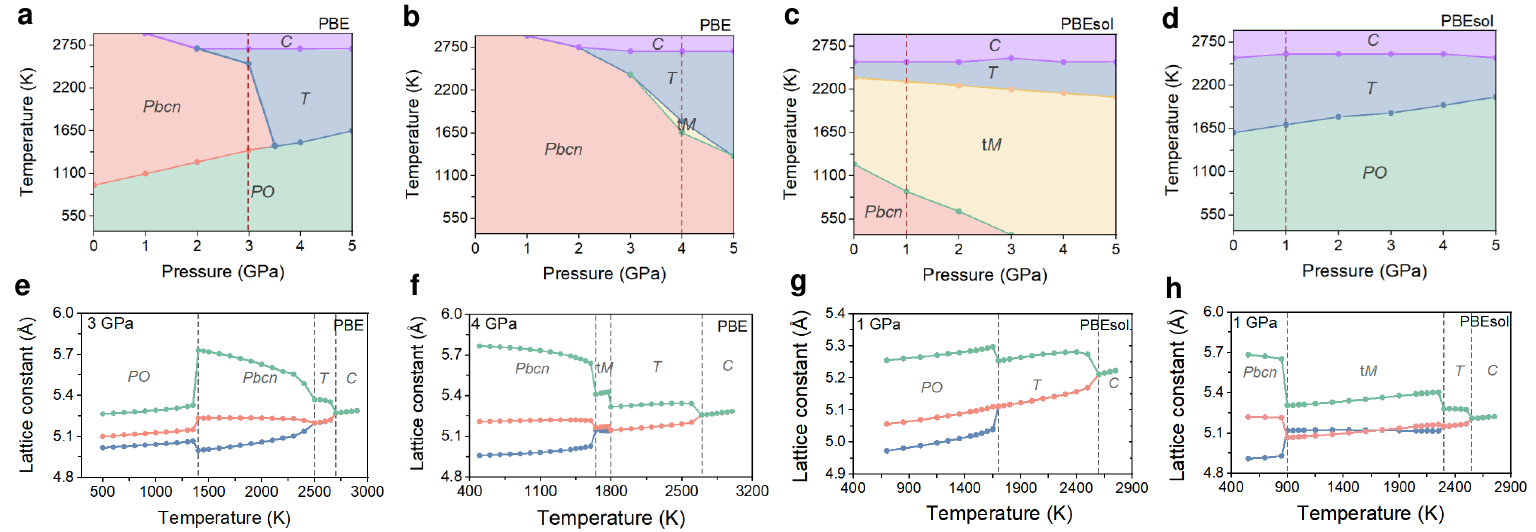}
\caption{Temperature-pressure phase diagrams of HfO$_2$ obtained from MD simulations using the DP model trained with PBE and PBEsol functionals. Simulations are initialized in (\textbf{a}) $PO$ and (\textbf{b}) $Pbcn$ using the PBE-based DP model, and in (\textbf{c}) $PO$ and (\textbf{d}) $Pbcn$ using the PBEsol-based DP model. The temperature-dependent lattice constants at representive pressures are shown in (\textbf{g}-\textbf{h}).}
\label{fig:phasediagram}
\end{figure}

\begin{figure}[t]
\includegraphics[width=0.5\textwidth]{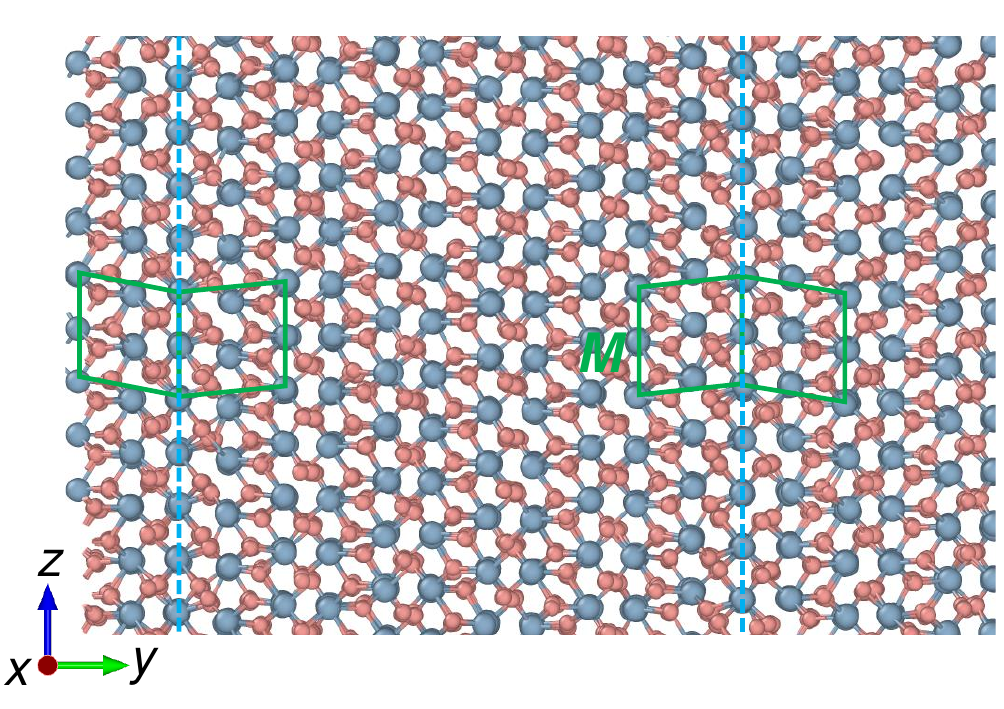}
\caption{Structure of t$M$ featuring twin boundaries between two $M$ domains.}
\label{fig:tM}
\end{figure}

We begin by mapping out the pressure-temperature phase diagrams using MD simulations based on the PBE-derived DP model. As shown in Fig.~\ref{fig:phasediagram}a, the system is initially in the $PO$ phase at 0~GPa, and after heating it transitions to the $Pbcn$ phase at 950 K. At elevated pressures (approximately 3.5 GPa and above), the $PO$ phase transitions directly to the nonpolar $T$ phase, with the transition temperature increasing with pressure. At higher temperatures (2700 K), it further transitions to the $C$ phase. The evolution of lattice constants with increasing temperature at 3.0~GPa is shown in Fig.~\ref{fig:phasediagram}e. The lattice constants exhibit an abrupt change during the $PO \rightarrow Pbcn$ transition, indicating a first-order phase transition. In contrast, the transitions $Pbcn \rightarrow T$ and $T \rightarrow C$ display relatively smooth variations in lattice constants, suggesting the possibility of second-order phase transitions. We further simulate the phase transitions by initializing the system in the $Pbcn$ phase, which is predicted by PBE to be more stable than the $PO$ phase. As illustrated in Fig.~\ref{fig:phasediagram}b, the temperature-driven phase transition sequence varies subtly at different pressures but mainly follows $Pbcn \rightarrow T \rightarrow C$ (see evolution of lattice constants in Fig.~\ref{fig:phasediagram}f). Notably, the transition temperature for $Pbcn \rightarrow T$ decreases with increasing pressure. The absence of the $PO$ phase in this phase diagram aligns with PBE-based QHA calculations, which consistently show that the $Pbcn$ phase has lower free energy than the $PO$ phase over a wide temperature range. We observe the emergence of an intermediate structure, denoted as t$M$, within the temperature range of 1650~K to 1750~K under pressures of 3 to 5~GPa. The t$M$ structure, illustrated in Fig.~\ref{fig:tM}, can be interpreted as the twinning of two $M$ domains, which has been observed experimentally~\cite{Du21p986}.

The phase diagrams obtained from MD simulations using the PBEsol-based DP model are presented in Fig.~\ref{fig:phasediagram}c-d. As discussed earlier, the PBEsol-based QHA calculations predict that the $PO$ phase remains consistently more stable than the $Pbcn$ phase across all temperatures, with a stability crossover observed between the $PO$ and $T$ phases at elevated temperatures. This prediction is consistent with the results shown in Fig.~\ref{fig:phasediagram}c, where MD simulations initialized in the $PO$ phase reveal that, as the temperature increases, the system transitions into the $T$ phase, which then transforms into the $C$ phase. Furthermore, the transition temperature exhibits a positive correlation with increasing pressure, indicating that higher pressures further stabilize the $PO$ phase prior to the $PO \rightarrow T$ transition.

Figure~\ref{fig:phasediagram}d illustrates the phase transition behavior when the system is initialized in the $Pbcn$ phase, which PBEsol predicts to be less stable than $PO$. The system first transitions into the t$M$ phase before further transforming into the $T$ phase at higher temperatures, with temperature-dependent lattice constants at 1~GPa reported in Fig.~\ref{fig:phasediagram}h. As pressure increases, the transition temperature from $Pbcn$ to t$M$ gradually decreases. At pressures exceeding 3~GPa, the $Pbcn$ phase becomes unstable and directly transitions into the t$M$ phase. In contrast to the PBE-based phase diagram shown in Fig.~\ref{fig:phasediagram}b, the PBEsol-based DP model reveals a broader temperature-pressure stability region for the t$M$ phase.

\subsection{PBE or PBEsol?}

For a given XC functional, the phase transition sequences predicted by QHA calculations consistently align with those obtained from DPMD simulations when the DP model is trained using the same functional. This raises a critical yet challenging question: which XC functional most accurately represents the ``ground truth" of the system, particularly in the comparison between PBE and PBEsol? If we follow the conventional assumption that nonlocal hybrid functionals such as HSE06 (situated on the fourth rung of Jacob’s ladder of density functional approximations) are inherently more accurate, PBE might be favored over PBEsol, as PBE reproduces the energy hierarchy predicted by HSE06. However, this assumption comes with an important caveat: the results of HSE06 are sensitive to the choice of the mixing parameter $\alpha$. The temperature-driven phase transitions in HfO$_2$ also exhibit remarkable complexity, depending sensitively on the initial phase and mechanical boundary conditions (\eg, pressure). For instance, a PBEsol-trained DP model predicts a phase transition sequence of $PO\rightarrow T \rightarrow C$ at 0~GPa, agreeing well with experimental observations. However, it is important to note that experimentally synthesized ferroelectric hafnia thin films are often polycrystalline and host multiple phases. This makes it challenging to definitively rule out the possible appearance of the $Pbcn$ phase during heating, as predicted by the PBE-based DP model. Adding further complexity to this already intricate issue, we find that the PBE-based DP model also predicts the $PO \rightarrow T \rightarrow C$ phase transition in HfO$_{2-x}$ with oxygen vacancies.

Given these complexities, we offer a cautious conclusion: The phase behavior of HfO$_2$ is governed by intricate thermodynamic and kinetic factors, and a definitive resolution will require synergistic advances in theory (\eg, higher-accuracy methods like quantum Monte Carlo)~\cite{Chimata19p075005,Shin18p075001} and experiment (\eg, in situ characterization of phase evolution under controlled conditions). Until then, the choice between PBE and PBEsol must balance empirical agreement with acknowledgment of their inherent approximations.

\subsection{Strain modulated switching pathway}

\begin{figure}[t]
\includegraphics[width=0.8\textwidth]{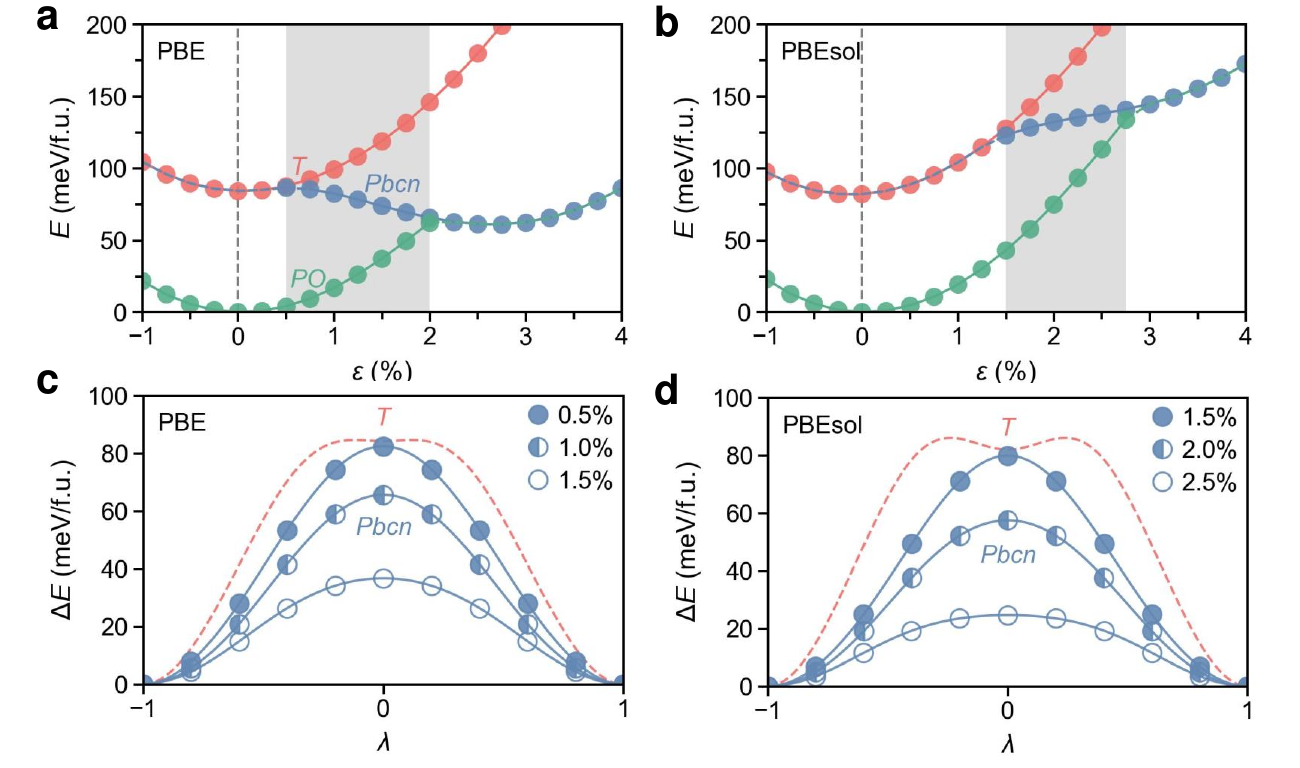}
\caption{Hydrostatic strain-dependent stability of the $PO$, $Pbcn$, and $PO$ phases computed using (\textbf{a}) PBE and (\textbf{c}) PBEsol. The lattice constants of the ground-state $PO$ phase serve as the zero-strain ($\epsilon=0$) reference. The shaded region indicates the strain range where the $Pbcn$ phase is lower in energy than the $T$ phase. The energy evolution along the minimum energy path for polarization reversal in the $PO$ phase is shown at representative strain values, as determined using the NEB method with (\textbf{b}) PBE and (\textbf{d}) PBEsol.}
\label{fig:strain1}
\end{figure}

Although PBE and PBEsol predict opposite relative stabilities between the $PO$ and $Pbcn$ phases, our model calculations tracing the energy evolution for $PO \rightarrow PO^*$ suggest that the two functionals yield comparable results. Specifically, under fixed mechanical boundary conditions, the energy evolution in response to atomic distortions appears to be relatively insensitive to the choice of XC functional. To demonstrate this, we compare the dependence of the relative stability of the $PO$, $Pbcn$, and $T$ phases on hydrostatic strain using the PBE and PBEsol functionals. The ground-state structure of the $PO$ phase is taken as the reference zero-strain state. For a given hydrostatic strain ($\epsilon$), the lattice constants along all three Cartesian axes are uniformly scaled by the same percentage, followed by relaxation of the atomic positions. As illustrated in Fig.~\ref{fig:strain1}a, our PBE calculations reveal that the $Pbcn$ phase is unstable when the lattice constants are fixed to the ground-state values of the $PO$ phase ($\epsilon=0$), with the structure spontaneously evolving into the $T$ phase during atomic relaxation. This behavior persists under compressive strains ($\epsilon<0$). With increasing tensile strain, the $Pbcn$ phase becomes more stable than the $T$ phase, while the energy difference between the $Pbcn$ and $PO$ phases gradually decreases. Beyond a critical tensile strain of $\epsilon > 2$\%, the $PO$ phase becomes unstable, spontaneously transforming into the $Pbcn$ phase during relaxation. Notably, as shown Fig.~\ref{fig:strain1}b, PBEsol predicts a qualitatively similar energy evolution for the three phases as a function of strain, albeit with a shifted strain range (shaded region) in which the $Pbcn$ phase is less stable than the $PO$ phase but more stable than the $T$ phase.

Previous theoretical studies have revealed that strain can alter the polarization switching mechanism. Specifically, the minimum energy path (MEP) may involve either the $T$ or $Pbcn$ phase as an intermediate state, depending on the strain conditions ~\cite{Zhou22peadd5953}. The hydrostatic strain-induced changes in the relative stability of the $PO$, $T$, and $Pbcn$ phases suggest that applying tensile strain can effectively lower the switching barrier by involving the low-energy $Pbcn$ phase during polarization reversal. Indeed, as shown in the MEPs determined using NEB (Fig.~\ref{fig:strain1}c-d), both PBE and PBEsol predict the emergence of an intermediate $Pbcn$ phase when the strain exceeds a critical threshold (0.5\% for PBE and 1.5\% for PBEsol). This leads to a substantial reduction of the switching barrier ($\Delta U$) with increasing tensile strain. For a given strain, we find that the $\Delta U$ value obtained using the NEB method closely matches the energy difference between the $Pbcn$ (or $T$) and $PO$ phases.

\begin{figure}[t]
\includegraphics[width=0.8\textwidth]{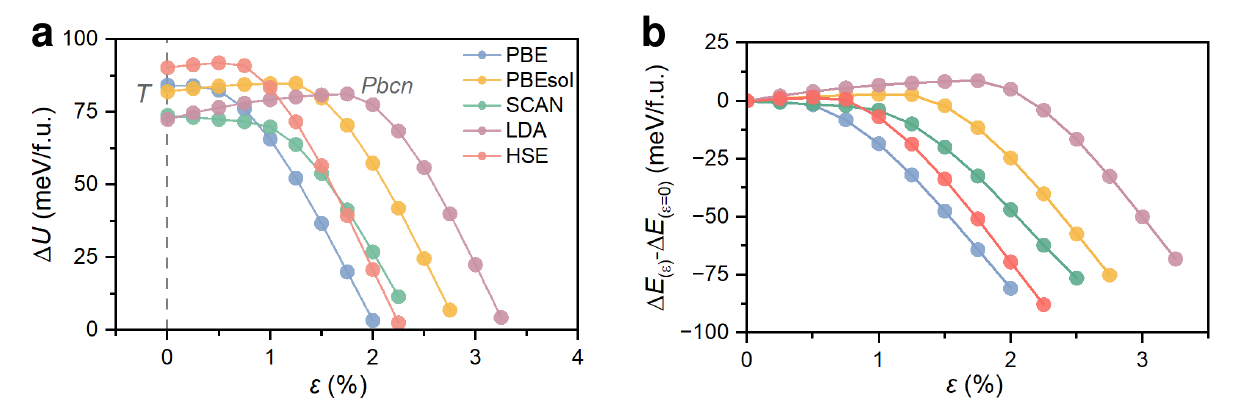}
\caption{(\textbf{a}) Comparison of strain-dependent switching barriers ($\Delta U$) predicted by different XC functionals. (\textbf{b}) Comparison of the change in the switching barrier relative to the strain-zero value, $\Delta U_{(\epsilon)}-\Delta U_{(\epsilon=0)}$.}
\label{fig:strain3}
\end{figure}

Figure~\ref{fig:strain3}a compares the strain-dependent $\Delta U$ across different XC functionals, where $\Delta U$ is approximated by the energy difference between the $PO$ phase and the intermediate $T$ or $Pbcn$ phase. All functionals exhibit a similar trend: the barrier remains nearly constant under low tensile strain but drops sharply beyond a critical threshold when the intermediate state transitions to the $Pbcn$ phase during switching. The main distinction among the tested functionals is the strain range over which the $Pbcn$ phase is relatively stable between the $PO$ and $T$ phases. Figure~\ref{fig:strain3}b illustrates the variation in $\Delta U$ relative to its zero-strain value. Notably, PBE predicts the lowest critical strain, while LDA estimates the highest. Despite these differences, all functionals yield comparable slopes in the regime where the barrier decreases. These results provide reassurance that prior theoretical studies on ferroelectric HfO$_2$ switching barriers using various XC functionals maintain general consistency, as lattice parameters are typically constrained to the ground-state values of the $PO$ phase.

\subsection{Proper or improper ferroelectricity?}

\begin{figure}[t]
\includegraphics[width=0.7\textwidth]{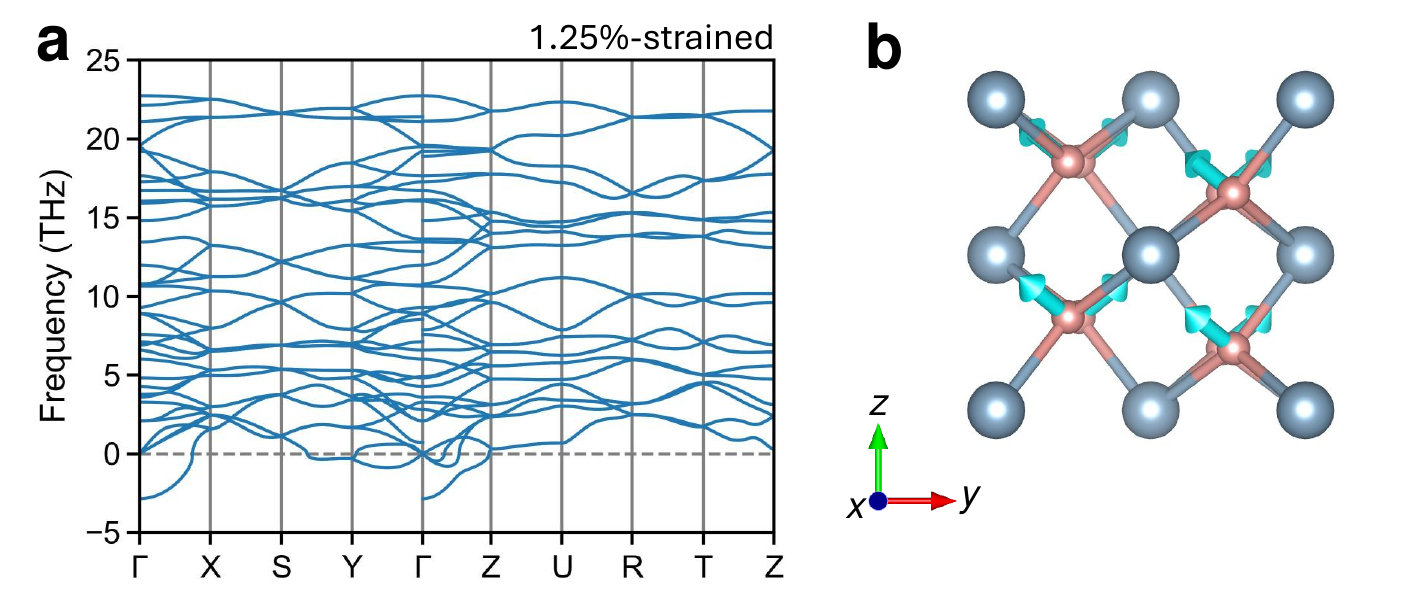}
\caption{(\textbf{a}) Phonon spectrum of the $Pbcn$ phase under a hydrostatic tensile strain of 1.25\%. (\textbf{b}) Schematic illustration of the $\Gamma$-point vibrational mode with an imaginary frequency. Blue arrows on the oxygen atoms represent their displacement directions.}
\label{fig:phostrain}
\end{figure}

The proper or improper nature of ferroelectric hafnia has been the subject of long-standing debate. Depending on the choice of the parent nonpolar phase (switching intermediate)~\cite{Zhou22peadd5953,Raeliarijaona23p094109,Aramberri23p95}, HfO$_2$ can be classified as either a proper or improper ferroelectric. We propose that this classification is strain-dependent. For strain-free or compressively strained HfO$_2$, polarization switching involves a nonpolar $T$ phase, suggesting it is better to recognize the $PO$ phase as an improper ferroelectric, as confirmed by the triple-well energy landscape (see red line in Fig.~\ref{fig:strain1}c). In contrast, under tensile strain within a certain range (denoted by the grey shaded area in Fig.~\ref{fig:strain1}a), the switching process involves the antipolar $Pbcn$ phase. As shown in Fig.~\ref{fig:phostrain}a, we find that under a 1.25\% tensile strain, the $Pbcn$ phase exhibits an imaginary frequency at the Gamma point, corresponding to the emergence of a polar mode (Fig.~\ref{fig:phostrain}b). The onset of this polar mode in the $Pbcn$ phase drives a transition to the $PO$ phase, thereby resulting in the formation of a structure with distinctly alternating polar and nonpolar oxygen atoms. Thus, it is appropriate to classify the $PO$  phase as a proper ferroelectric, consistent with the double-well energy landscape (see blue lines in Fig.~\ref{fig:strain1}c).

\subsection{Domain wall energy}

\begin{figure}[t]
\includegraphics[width=0.8\textwidth]{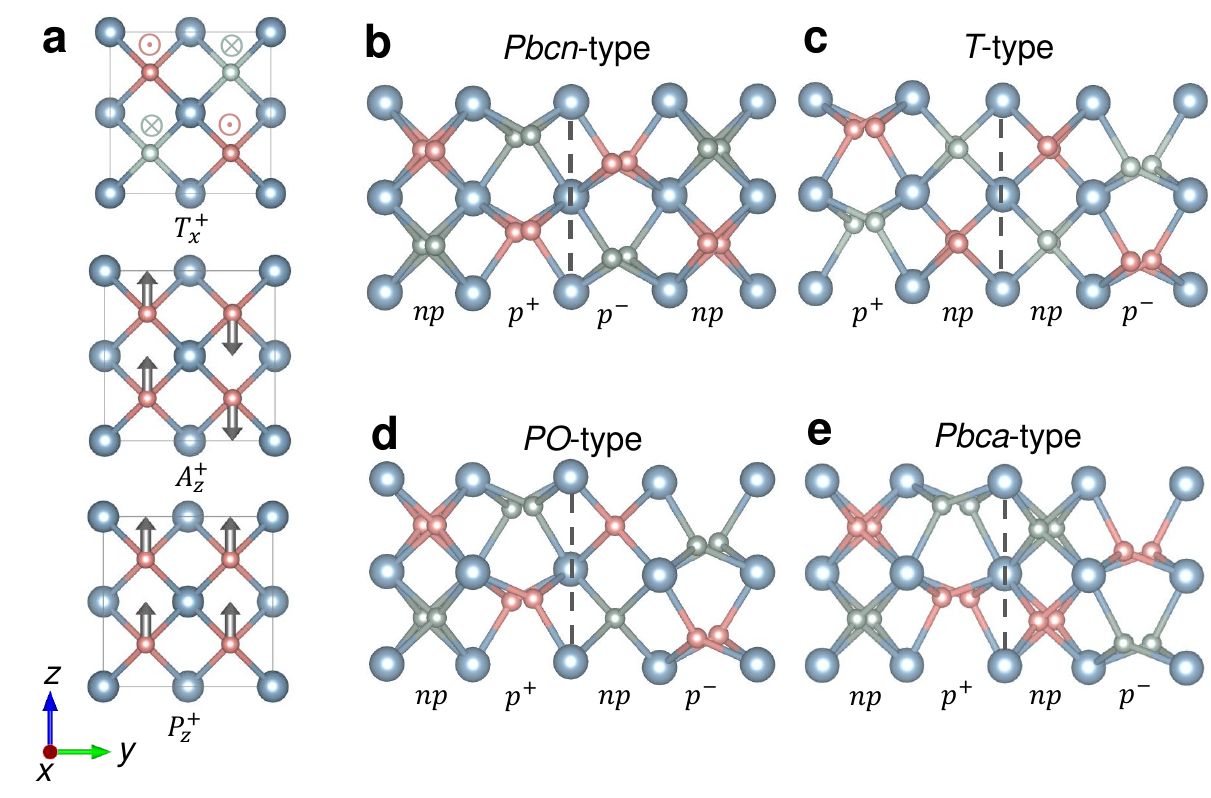}
\caption{Four distinct types of 180\degree~domain walls characterized by lattice mode analysis. (\textbf{a}) Lattice mode analysis for the $Pca2_1$ phase. The tetragonal mode ($T_x^+$) involves oxygen atoms displacing in an antiparallel pattern along the $x$ direction, with outward and inward displacements shown in red and green, respectively. The antipolar mode ($A_z^+$) features oxygen displacements along the $z$-direction in an antipolar pattern, while the polar mode ($P_z^+$) involves all oxygen atoms undergoing polar displacements along the $z$-direction. (\textbf{b}) $Pbcn$-type and (\textbf{c}) $T$-type domain walls, characterized by polar ($p$) and nonpolar ($np$) oxygen atoms on both sides of the wall, respectively. (\textbf{d}) $PO$-type (\textbf{e}) and $Pbca$-type domain walls, where $p$ and $np$ oxygen atoms are located on opposite sides of the wall. These two wall types are distinguished by the behavior of the $T_x$ mode: in the $PO$-type wall, alternating red (outward) and green (inward) oxygen atoms appear across the interface, preserving the $T_x$ mode sign. In contrast, the $Pbca$-type wall exhibits a reversal of the $T_x$ mode, with neighboring oxygen atoms at the interface sharing the same color.}
\label{fig:DWstructure}
\end{figure}

Domain walls are critical topological defects in ferroelectrics and often play a pivotal role in polarization switching. Therefore, it is important to investigate the dependence of domain wall energy in HfO$_2$ on the choice of XC density functional, especially since certain types of domain walls locally adopt a $Pbcn$-like structure. In the $PO$ phase of HfO$_2$, a diverse array of domain wall configurations can form due to its unique structural characteristics. The unit cell of $PO$ HfO$_2$ consists of alternating layers of fourfold-coordinated oxygen ions (often referred to as nonpolar oxygen, $np$) and a layer of threefold-coordinated oxygen ions (polar oxygen, $p$). This layered configuration can be understood as the result of successive condensation of three lattice modes: $T_x$, $A_z$, and $P_z$. As illustrated in Fig.~\ref{fig:DWstructure}a, in a cubic reference phase, the $T_x$ mode is characterized by antiparallel $x$-displacements of neighboring oxygen atoms, the $A_z$ mode involves antiparallel $z$-displacements, and the $P_z$ mode corresponds to parallel $z$-displacements of all oxygen atoms. Starting from the cubic phase, successive lattice distortions via the $T_x$, $A_z$, and $P_z$ modes lead to the formation of the tetragonal $P4_2/nmc$, antipolar $Pbcn$, and polar $Pca2_1$ phases, respectively. As a result, the unit cell of the $PO$ phase can be described by the mode vector $(T_x, A_z, P_z)$.

There are four main types of 180$\degree$ domain walls that separates oppositely polarized $PO$ domains in HfO$_2$, distinguished by how $T_x$, $A_z$, and $P_z$ modes evolve across the interface. As shown in Fig.~\ref{fig:DWstructure}b, the $Pbcn$-type wall separates domains with mode vectors $(T_x^+, A_z^-, P_z^+)$ and $(T_x^+, A_z^-, P_z^-)$. The unit cell at the wall adopts a $Pbcn$-like structure, characterized by antiparallel polar oxygen atoms. The $T$-type domain wall, illustrated in Fig.~\ref{fig:DWstructure}c, forms between regions with $(T_x^+, A_z^+, P_z^+)$ and $(T_x^+, A_z^+, P_z^-)$ mode vectors. This results in a $T$-like configuration, where nonpolar oxygen atoms are present on both sides of the wall. In contrast, the $PO$-type and $Pbca$-type domain walls feature interfaces with neighboring polar and nonpolar oxygen atoms. Specifically, the $PO$-type domain wall (Fig.~\ref{fig:DWstructure}d) separates $(T_x^+, A_z^-, P_z^+)$ and $(T_x^+, A_z^+, P_z^-)$, featuring a $PO$-like interface with conserving $T_x$ mode across the wall. While the $Pbca$-type domain wall (Fig.~\ref{fig:DWstructure}e) separates $(T_x^+, A_z^-, P_z^+)$ and $(T_x^-, A_z^+, P_z^-)$, adopting a $Pbca$-like structure. Unlike the $PO$-type wall, it exhibits a reversal of the $T_x$ mode across the interface.

We note that due to the periodic boundary conditions employed in DFT calculations, using a supercell consisting of two oppositely polarized domains with an even number of $PO$ unit cells (\eg, a $1\times8\times1$ supercell containing two $1\times4\times1$ domains) will inevitably result in the simultaneous presence of both $Pbcn$-type and $T$-type domain walls. To ensure the presence of two identical domain walls within the supercell, we instead use a $1\times6\times1$ supercell. The domain wall energy is calculated by first determining the total energy of the supercell containing two identical domain walls, then subtracting the total energy of the ground-state $Pca2_1$ phase. The difference is then divided by 2 to obtain the energy of a single domain wall.
 
Table~\ref{tab:dw} summarizes the calculated domain wall energies ($\sigma_{\mathrm{DW}}$) of HfO$_2$ at 0 K using various XC functionals. All functionals consistently predict that the $Pbca$-type wall is the most stable, with the energy of the $PO$-type wall always falling between those of the $Pbcn$-type and $T$-type walls. However, the relative energies of the $Pbcn$-type and $T$-type walls depend notably on the choice of functional. PBE predicts a lower energy for the $Pbcn$-type wall compared to the $T$-type wall, whereas the other three functionals yield the opposite trend. This behavior is consistent with unit-cell calculations, where PBE predicts the $Pbcn$ phase to be lower in energy than the $T$ phase. Overall, all functionals predict comparable domain wall energies.

\section{CONCLUSION}

As an antipolar phase, the $Pbcn$ phase of HfO$_2$ has been suggested to play an important role in both phase transitions and polarization switching mechanisms in ferroelectric hafnia. Despite extensive theoretical studies on HfO$_2$, it is surprising that the strong dependence of the $Pbcn$ phase's thermodynamic stability on the choice of XC density functional, relative to the $Pca2_1$ and $P4_2/nmc$ phases, has largely gone unnoticed. Our systematic investigations reveal that PBE and HSE06, using the default mixing parameter ($\alpha = 0.25$), predict $Pbcn$ to be more stable than $Pca2_1$, whereas LDA, PBEsol, and SCAN all yield the opposite energy ordering. Finite-temperature molecular dynamics simulations using machine learning force fields trained with PBE and PBEsol further reveal complex temperature-pressure phase diagrams for both the $Pca2_1$ and $Pbcn$ phases. Given the intricate nature of phase transitions among hafnia polymorphs, we strongly encourage future theoretical studies using higher-accuracy methods, such as quantum Monte Carlo, in combination with detailed experimental investigations, including in situ characterization of phase evolution under controlled conditions.

Reassuringly, if we assume mechanical boundary conditions are dictated by the ground-state lattice constants of the $Pca2_1$ phase, different density functionals predict similar trends in strain-dependent switching barriers and domain wall energies. However, caution is needed regarding the exact critical strain value, as it exhibits strong functional dependence. Overall, our findings highlight the significant impact of density functional choice on the stability and phase behavior of HfO$_2$, underscoring the need for both theoretical and experimental efforts to refine our understanding of its polymorphic transitions.

\vspace{1cm}

{\bf{Acknowledgments}} D.F., T.Z. and S.L. acknowledge the support from National Key R\&D Program of China (Grant No.~2021YFA1202100), National Natural Science Foundation of China (Grants No.~12361141821, No.~12074319), and Westlake Education Foundation. The computational resource is provided by Westlake HPC Center.

{\bf{Author Contributions}} S.L. and T.Z. conceived the idea and designed the project. D.F. wrote a major part of the paper, S.L. and T.Z. revised the paper. D.F. and T.Z. contributed equally to this work. All authors reviewed and edited the manuscript.

{\bf{Competing Interests}} The authors declare no competing financial or non-financial interests.

{\bf{Data Availability}} The data that support the findings of this study are included in this article and are available from the corresponding author upon reasonable request.

\clearpage

\begin{table}[ptb]
\caption{Dependence of lattice parameters and relative energies of different phases of HfO$_2$ on the choice of exchange-correlation functional, using the $P2_1/c$ ($M$) phase as the energy reference.}
\begin{ruledtabular}
\renewcommand{\tabcolsep}{0.005pc}
\begin{tabular}{ccccccc@{\hskip 0.6em}|cccc@{\hskip 0.6em}|c}
& &\multicolumn{5}{c}{VASP} & \multicolumn{4}{c}{PWmat}&Expt.\\\cline{3-12}
Phase&& LDA & PBE & PBEsol & SCAN & HSE & LDA & PBE & PBEsol & HSE & \\\hline
\multirow{4}{*}{$M$} &$a$ (\AA)&5.03&5.14&5.08&5.07&5.10&4.95&5.08& 5.03&5.05&5.12 \cite{Ruh70p126} \\ 
&$b$ (\AA)&5.12&5.19&5.16&5.14&5.14&5.04&5.13&5.10&5.10&5.17 \cite{Ruh70p126}\\ 
&$c$ (\AA)&5.20&5.32&5.25&5.26&5.27&5.11&5.26&5.13&5.24&5.30 \cite{Ruh70p126} \\
&$E$ (meV/f.u.) &0.0&0.0&0.0&0.0&0.0&0.0&0.0&0.0&0.0 \\ \hline
\multirow{4}{*}{$Pbcn$} &$a$ (\AA)&4.86&4.91&4.88&4.86&4.87&4.78&4.85&4.83&4.82 \\ 
&$b$ (\AA)&5.13&5.25&5.19&5.17&5.20&5.05&5.18&5.13&5.14 \\ 
&$c$ (\AA)&5.62&5.78&5.70&5.70&5.71&5.53&5.70&5.62&5.64 \\ 
&$E$ (meV/f.u.) &175.6&46.2&126.7&76.1&65.6&161.5&24.3&103.8&30.9 \\ \hline
\multirow{4}{*}{$PO$} &$a$ (\AA)&4.96&5.05&5.00&4.99&5.00&4.88&4.99&4.95&4.96&5.00 \cite{Hoffmann15p072006} \\ 
&$b$ (\AA)&4.98&5.08&5.02&5.02&5.03&4.90&5.02&4.97&5.00&5.06 \cite{Hoffmann15p072006} \\
&$c$ (\AA)&5.16&5.27&5.20&5.21&5.21&5.08&5.21&5.16&5.17&5.23 \cite{Hoffmann15p072006} \\ 
&$E$ (meV/f.u.) &52.9&84.3&64.3&66.1&78.0&61.9 &103.1&77.8&77.5\\ \hline
\multirow{4}{*}{$T$}&$a$ (\AA)&4.98&5.08&5.02&5.02&5.03&4.90&5.01&4.98&4.99&5.06 \cite{MacLaren09pG103}\\ 
&$b$ (\AA)&4.98&5.08&5.02&5.02&5.03&4.90&5.01&4.98&4.99&5.06 \cite{MacLaren09pG103} \\ 
&$c$ (\AA)&5.07&5.23&5.13&5.18&5.15&5.00&5.17&5.10&5.12&5.20 \cite{MacLaren09pG103}\\
&$E$ (meV/f.u.)&115.6 &166.2&139.6&137.9&163.8&118.5&117.1&143.9&145.7 \\ 
\end{tabular}
\end{ruledtabular}
\label{tab:phase}
\end{table}

\clearpage

\clearpage

\clearpage

\clearpage

\clearpage

\clearpage

\clearpage

\clearpage

\clearpage

\clearpage

\clearpage

\clearpage

\begin{table}[ptb]
\caption{Domain wall energy ($\sigma_{\mathrm{DW}}$, in unit of meV/\AA$^2$) estimated with four different XC functionals.}
\begin{ruledtabular}
\renewcommand{\tabcolsep}{0.006pc}
\begin{tabular}{ccccc}
& $Pbcn$-type & $T$-type & $PO$-type & $Pbca$-type \\ \hline
PBE & 18.4 & 20.6 & 18.8 & $-$1.1 \\ \hline
PBEsol & 27.6 & 24.1 & 25.7 & $-$2.3 \\ \hline
SCAN & 22.5 & 21.7 & 22.4 & $-$1.4 \\ \hline
LDA & 32.1 & 24.8 & 27.8 & $-$2.6 \\ 
\end{tabular}
\end{ruledtabular}
\label{tab:dw}
\end{table}

\clearpage

\bibliography{SL}

\end{document}